\documentstyle[multicol,aps,prl,epsf,psfig]{revtex}
\begin{document}
\def\be{\begin{equation}}
\def\ee{\end{equation}}
\def\lag{\langle}
\def\rag{\rangle}
\def\ep{\epsilon}
\def\aep{\vert \epsilon \vert}
\def\betak{{\beta^{\rm KPZ}}}
\def\chik{{\chi^{\rm KPZ}}}
\def\zk{{z^{\rm KPZ}}}
\def\chiis{{\chi^{\rm isot}}}
\title{Delocalisation transition of a rough adsorption-reaction
interface} 
\author{H\"useyin Kaya $^{1,3}$,
Alkan Kabak\c c\i o\u glu $^{1,2}$,
and Ay\c se Erzan$^{1,3}$}
\address{$^1$  G\"ursey Institute, P. O. Box 6, \c
Cengelk\"oy 81220, Istanbul, Turkey}
\address{$^2$ Department of Physics and Center for Material
Sciences and Engineering, 
\\Massachusetts Institute of
Technology, 
 Cambridge, Massachusetts 02139, USA}
\address{$^3$  Department of Physics, Faculty of  Sciences
and
Letters\\
Istanbul Technical University, Maslak 80626, Istanbul, Turkey }
\maketitle
\begin{abstract}
We introduce a new kinetic interface model suitable for
simulating adsorption-reaction processes which take place
preferentially at surface defects such as steps and vacancies.
As the average interface velocity is taken to zero, the self-
affine interface with Kardar-Parisi-Zhang like scaling behaviour
undergoes a delocalization transition with critical exponents
that fall into a novel universality class. As the critical point
is approached, the interface becomes a multi-valued, multiply
connected self-similar fractal set.  The scaling behaviour and
critical exponents of the relevant correlation functions are
determined from Monte Carlo simulations and scaling arguments.

PACS Numbers: 68.35.Ct, 82.65.Jv, 05.70.Ln, 68.35.Rh
\end{abstract}
\begin{multicols}{2}
Kinetically roughened interfaces display a rich phenomenology,
have deep connections with fields as diverse as self-organized
criticality, spin-glasses and complex pattern formation, and lend
themselves to modelling various systems with practical
applications, ranging from heterogenous catalysis to
geomorphology~\cite{Tim,Derrida,KS,Meakin}.  The huge amount of
numerical and analytical effort that has recently been invested
in them has revealed that they obey universal scaling relations,
which fall into one of a few universality classes.
In this letter we would like to present a kinetic interface model
which exhibits an anisotropic to isotropic phase transition with
novel scaling behaviour at the delocalisation critical point.

Reaction fronts formed by $A+B \to \emptyset$ reactions in
heterogeneous
systems where the reaction takes place on a two dimensional
substrate~\cite{Ziff} are often confined to a narrow ``reactive
zone'' especially if the reactants are either initially
segregated or become segregated due to reaction
kinetics~\cite{Cardy,Galfi,Kaya1,Kaya2}.
Our present model is motivated by recent
findings~\cite{Wandelt,Somorjai} of high reaction rates and
strong bonding at surface defects like steps and vacancies in
studies of heterogeneous catalysis, a burgeoning new field in
surface science.  

\begin{figure}
\begin{center}
\leavevmode
\psfig{figure=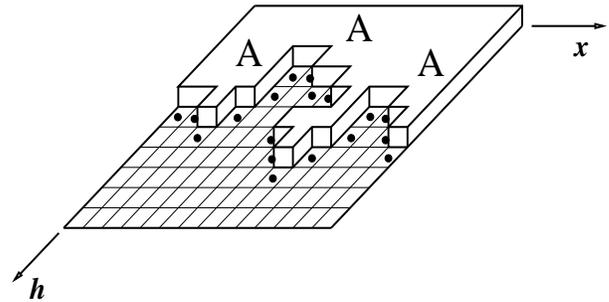,width=8cm,height=4cm,angle=0}
\end{center}
\narrowtext
\caption{ A terrace of $A$ particles, with the interface sites,
indicated
by dots, neighboring
the step.}
\label{step}
\end{figure}

We consider an idealised surface with  only one step, terminating
a terrace made up of $A$ particles (Fig.\ref{step}).  The surface
is
exposed to
two kinds of incoming particles, $A$ and $B$, which are allowed
to adsorb at first contact, and only on sites adjacent to the
step, which we will call ``interface sites.'' The adsorption of
$A$ particles makes the interface advance.  The adsorbing $B$
particles, on the other hand, immediately react with an $A$
neighbor to form a product which leaves the surface.  This 
eats into the step, making the interface recede. We investigate
the effect of changing the rate of injection~\cite{Richardson}
of the two reactants.
We do not allow any reactions to take place with the substrate
atoms. 
We assume, for
simplicity, that the temperature is low enough so that no surface
restructuring occurs; the bonding to the interface sites is
sufficiently strong~\cite{Memmelt} for diffusion along the
interface to be prohibited.  The kinetics is therefore driven by
the adsorption and reaction steps and not by the transport of the
reactants.

\begin{figure}{\vspace*{-2cm}}
\begin{center}
\leavevmode
\psfig{figure=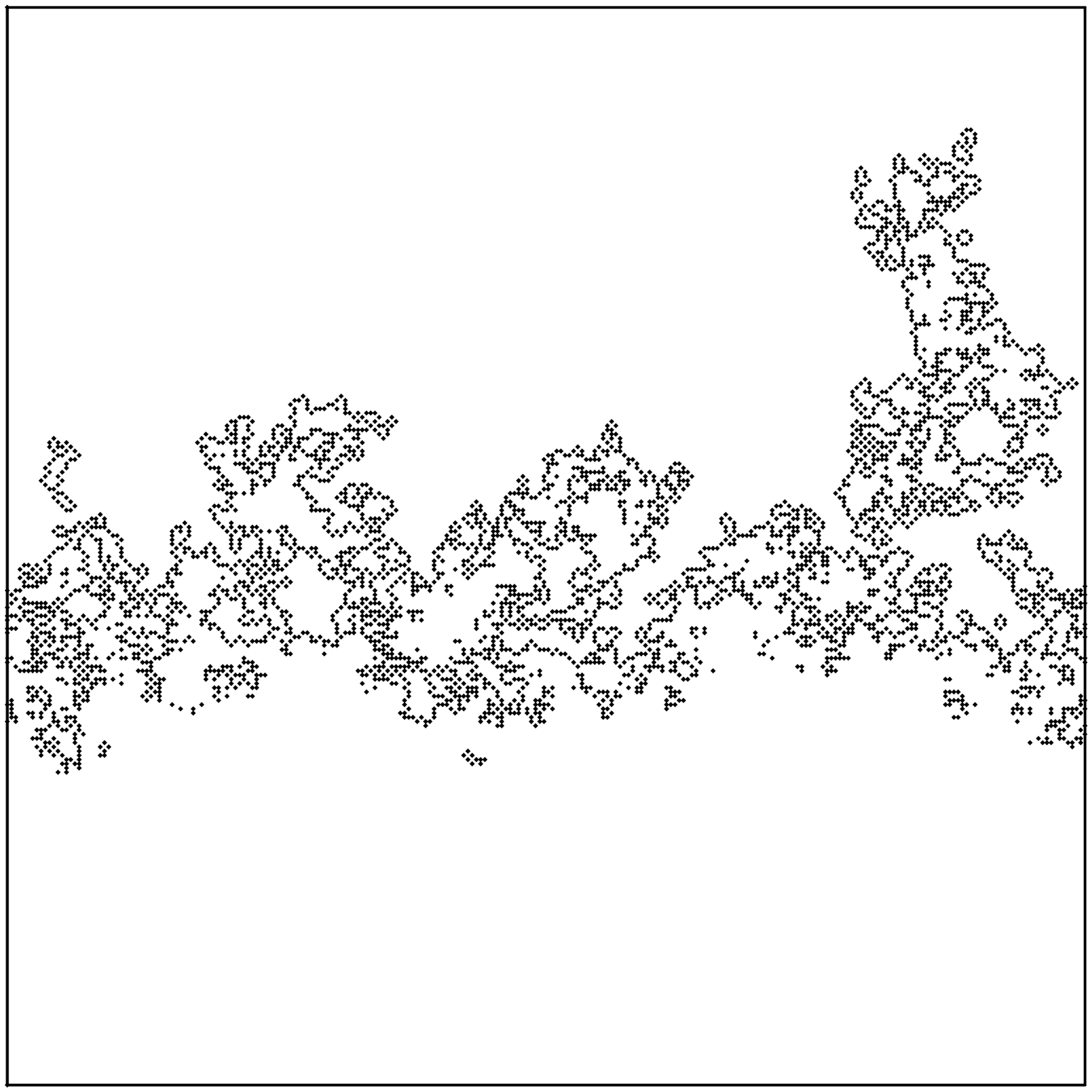,width=8cm,height=4cm,angle=0}
\end{center}
\narrowtext
\caption{ The interface at $\epsilon=0$, for early and later
stages
of growth. The configurations for early times resemble the
surface at larger values of $\epsilon$.}
\label{snapshot}
\end{figure}

The model is defined on an infinite strip of width $L$, on which
we impose periodic boundary conditions.  The interface is
initially a perfectly straight line located at $h=0$.  The system
is driven weakly so that at any instant only one particle of
either $A$ or $B$ type, with probabilities $p_A$ or $p_B=1-p_A$,
impinges on the interface.
As the interface
moves with a mean velocity equal to $\ep \equiv p_A-0.5$, it
roughens, and becomes multiply connected, shedding 
``islands'' or
``lakes''  in its wake. 
As $\aep \to 0$,  the growth direction is  completely
delocalised, the width of the interfacial region keeps on growing
indefinitely, and the interface breaks up into an isotropic
fractal (see Fig.\ref{snapshot}). For finite $L$, there may exist
more than one spanning
string of interface sites at $\epsilon=0$ ; this phenomenon is
similar to the formation of Liesegang 
bands~\cite{Liesegang}). 
It is the purpose of this letter to understand the
nature of this delocalization transition, to describe the
crossover behaviour and to characterise the self-similar reactive
region formed as $\aep \to 0$.

For many interface problems with a well defined growth direction,
such as the Eden~\cite{Eden} model, or the Edwards-Wilkinson
model~\cite{EW}, where the interface can be described with a
single-valued, self-affine curve. 
The scaling behaviour of the interface width 
may be conveniently summarized by the scaling
form~\cite{Vicsek}
\be 
w \sim t^\beta g(\ell/t^{1/z})\;\;,\label{Vic}
\ee where $g(u)\sim {\rm const}$ for $u<1$ and $\sim u^\chi$ for
$u\gg 1$;$z$ is the dynamical critical exponent,
and 
$\beta=\chi/z$ . Kardar, Parisi and Zhang~\cite{Kardar} have  
found the values $z=3/2$, $\chi
=1/2$ and $\beta=1/3$
for the stochastic
differential equations describing Eden growth in $d=1+1$. 
This set of critical
exponents characterizes a wide range of anisotropic growth
phenomena with annealed noise~\cite{Tim}, and where the local
velocity of the interface increases with the slope. In the limit
that the velocity goes to zero or 
is independent of the slope~\cite{KS}, one gets the
Edwards-Wilkinson model~\cite{EW}, which is exactly solvable in
$d=1+1$ dimensions and falls into another universality class,
characterized by $z=2$, $\chi=1/2$ and
$\beta=1/4$.
\begin{figure}
\begin{center}
\leavevmode
\psfig{figure=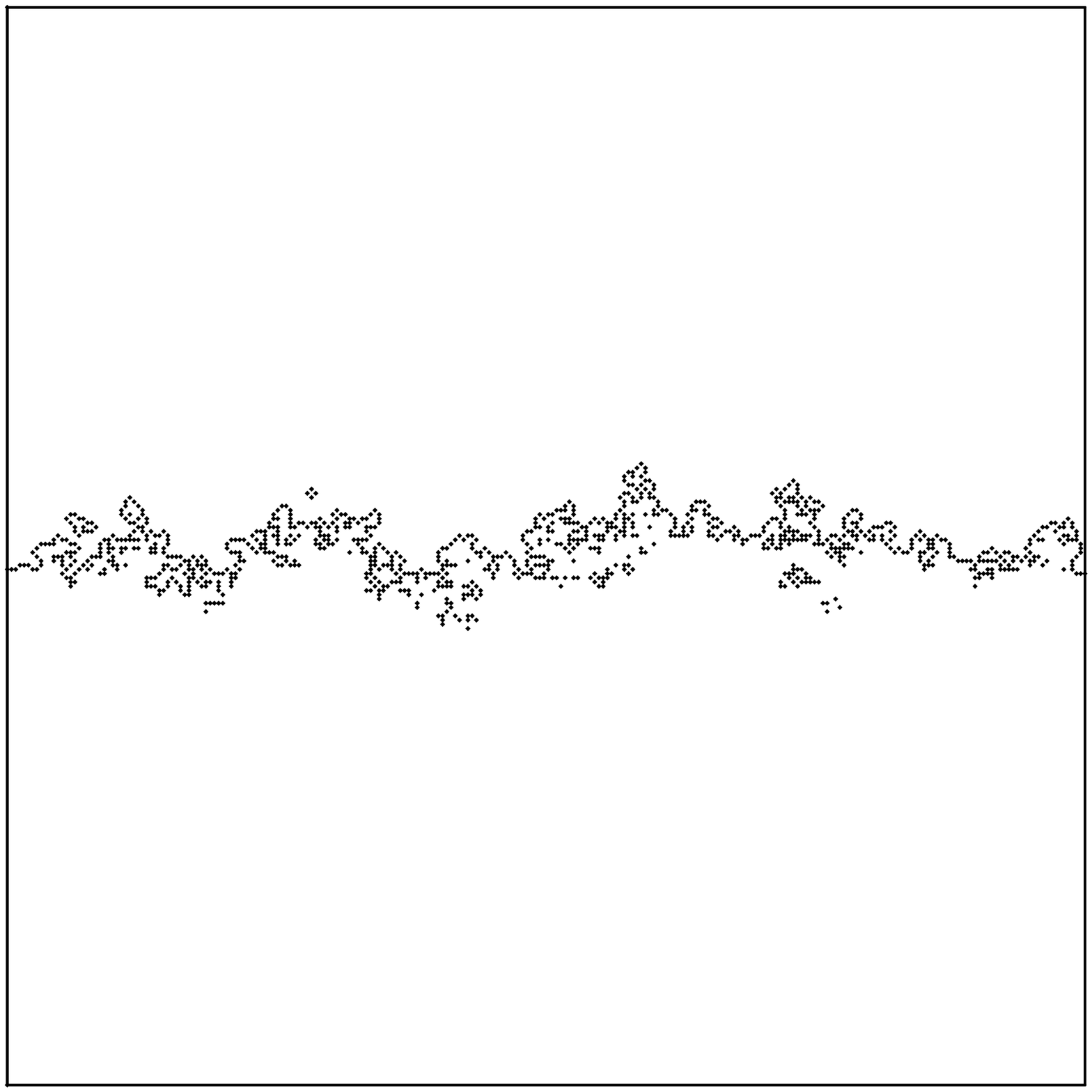,width=8cm,height=4cm,angle=0}
\end{center}
\end{figure}

\begin{figure}
\begin{center}
\leavevmode
\psfig{figure=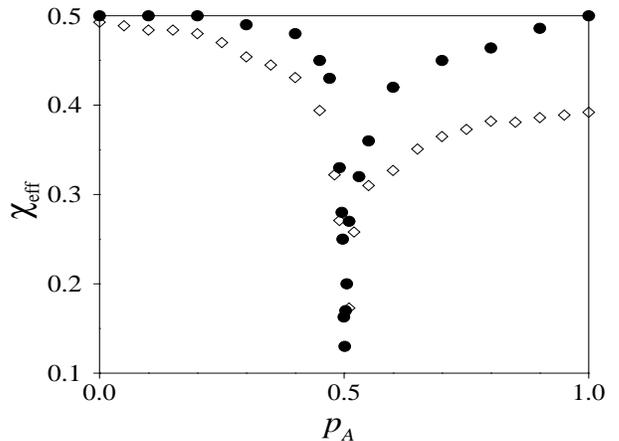,width=8cm,height=6cm,angle=0}
\end{center}
\narrowtext
\vspace*{-0.3cm}
\caption{The effective roughness exponent $\chi_{\rm eff}$, for
disconnected parts of the interface included in ($\diamond$) and
excluded
from ($\bullet$)
the analysis; $L\le 1024$. The error bars are comparable to the 
size 
of the symbols.}
\label{keff0l}
\end{figure}

Since our interface is typically multivalued we define
the width function within an interval of size $\ell$ as,
\be
w(\ell)^2 =\langle\,\, {1 \over N(\ell)} \sum_i^{N(\ell)}
(h_i-\overline{h(\ell)})^2\,\, \rangle \label{width}
\ee
where $h_i$ is the height of the $i$th
interfacial site, $i=1,\ldots,N(\ell)$, and
$\overline{h(\ell)}$ the mean position of the interface.
We find $\beta=1/3$ for early times. In the limit of $p_A=0$ or
$p_B=0$, i.e., for $\aep
\to  0.5$, our model is  equivalent to Eden~\cite{Eden}
growth, and indeed, along the singly connected part of the
interface
$\chi=1/2$ in the steady state.  However, effective 
roughness exponent ($\chi_{\rm eff}$) 
goes continuously to zero as $\aep \to 0$ as shown in 
Fig.\ref{keff0l}.  
The reason for this is that as we decrease
$\aep$, the surface becomes highly convoluted, 
with islands (or lakes) of all sizes, and therefore increasingly
multivalued. A competing length scale emerges in the system, the
``thickness'' of the interface, which we can measure by the 
variance of the height,
$$y(i)= \bigg( {1\over n_i} \sum_{j=1}^{n_i}
(h_{ij}-\overline{h_i})^2\bigg)^{1/2}\;\;,$$
with $n_i$  being the number
of interfacial sites  $\{h_{ij}\}$, above any point $i$ along the
horizontal axis.
The thickness obeys~\cite{son} a skewed--Gaussian
distribution. The average $y_L \equiv <y>_L$
and the second moment of this  distribution both diverge as
$\aep\to 0$
with a critical exponent $\nu =0.55\pm 0.05\,\simeq 1/2$, as 
$\sim \vert \ep \vert ^{-\nu}.$ 
The scaling form for $y_L$  is,
\be
y_L \sim t^{\tilde\beta} G(\aep^{-\nu}/t^{1/\zeta})\label{yL}
\ee 
where $G(v) \sim v$ for $v<1$ while for $v>1$, $G(v) \sim {\rm
const.}$ with $\tilde\beta =1/2$ and the (longitudinal) dynamical
critical exponent $\zeta$ obeys $\zeta=1/\tilde\beta$.
The thickness of just the singly connected part does not diverge
as $\aep \to 0$, so that the
disconnected 
parts make up almost all of the interfacial region at the
delocalization transition\cite{son}.

\begin{figure}
\begin{center}
\leavevmode
\psfig{figure=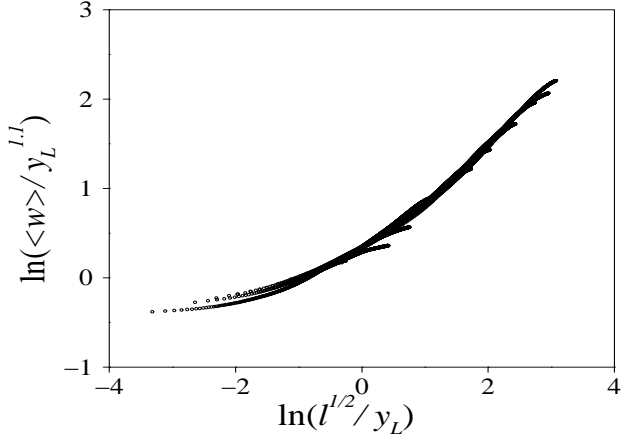,width=8cm,height=6cm,angle=0}
\end{center}
\narrowtext
\caption{The width normalized by a power of the thickness $y_L$
very near unity, displays KPZ
behaviour for $\ell^{1/2} /y_L >1$.  The different curves 
correspond to data taken at
$\ep=0.001$, 0.003, 0.005, 0.01, 0.05, 0.1, 0.2,
0.3, 0.4, 0.5. The deviations from the smooth collapse are due
to $\ell $
becoming comparable to the system size $L=1024$.}
\label{normyl1}
\end{figure}

Normalizing $w(\ell)$ in Eq.(\ref{width}) by $y_L^{1.1}$ yields
a collapse of the data for all $\ep$  as can be seen
from Fig.\ref{normyl1}.  We believe that the small deviation of
the power of $y_L$ from unity is due to insufficient statistics
for $y_L$ which converges extremely slowly as $\aep\to 0$ and may
safely be neglected, and we conclude 
\be
w(\ell) \sim  y_L \cases{ \ell^{1/2}/y_L & $\ell^{1/2}\gg y_L$\cr
  {\rm const.} & $\ell^{1/2} \ll y_L$.}
\label{yscale}\ee
Thus, $\chi_{\rm eff}$ goes to 
zero as the self-affine excursions of the interface are blurred
by the thickness of the interface  as
$y_L$ becomes greater than $ \ell^{1/2}$. 

If one considers  coarse--grained width functions,
either  by
taking the average height at any given point 
or  the maximum~\cite{Nolle} height for $\ep >0$
(minimum for $\ep <0$), one finds that they obey the scaling form
(\ref{Vic}) with KPZ exponents for $\ell\ll L$.

To get the local scaling picture, we focus on a single spanning
string in the interface and consider 
\begin{eqnarray}
C_x(l)&=&\langle\,\, (x(r+l)-x(r))^2\,\, \rangle ^{1/2}\;\;, \\
C_h(l)&=&\langle\,\, (h(r+l)-h(r))^2\,\, \rangle ^{1/2}\;\;.
\end{eqnarray}
where both $r$ and $l$ are the (``chemical") length measured
along the string, and  $x$ and $h$
are Cartesian coordinates of the interface
site. The scaling relations we have found from 
Fig.\ref{chcx} 
for these quantities are given below. 
In the transient regime ($l\gg t^{1/2}$), 
\be
C_x\sim\cases{ l & \,\,\, $y_L\ll l^{1/2}$ \cr
               l/t^{\psi} &\,\,\, $y_L\gg l^{1/2}$\cr }
\ee
and
\be C_h\sim t^{\beta}\ee
where $\psi=1/6$.

\begin{figure}
\begin{center}
\leavevmode
\psfig{figure=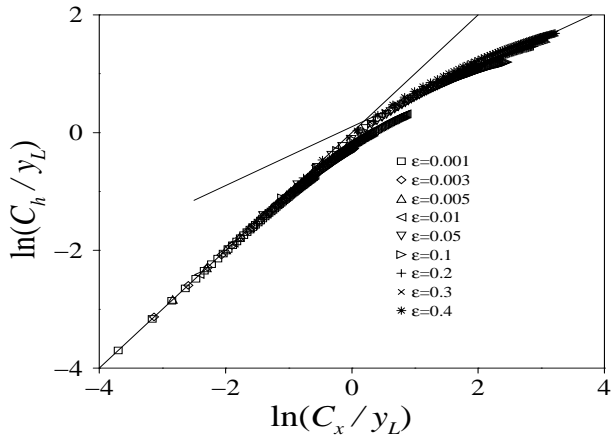,width=8cm,height=6cm,angle=0}
\end{center}
\narrowtext
\caption{ The vertical projection $C_h$ of sections of the
singly
connected part, plotted as a function of the horizontal
projection $C_x$.}
\label{coll}
\end{figure}

Note
the horizontal projection of a segment of fixed ``chemical
length" decreases  with $t$ as the surface
crumples with time in the critical ($\aep
\to 0$) region. In the steady state,
($l\ll t^{1/2}$), 
\be C_x \sim \cases{ l &$(y_L\ll l^{1/2})$ \cr
l^{\chi_{\rm isot}} &$(y_L\gg l^{1/2})$}\;\;.
\label{CxE2}
\ee
and
\be 
C_h\sim\cases{ l^{1/2} & $(y_L\ll l^{1/2})$\cr
l^{\chi_{\rm isot}} & $(y_L\gg l^{1/2})$. }
\label{Chc2}
\ee 
We see that for $y_L\gg l^{1/2}$, the interfacial region
becomes isotropic, 
with $C_x \sim
C_h$. This regime is characterized by an ``isotropic''
roughness exponent
$\chi_{\rm isot}=2/3$. In the opposite limit, $C_h\sim
C_x^{1/2}$, as expected for the self-affine Eden surface.  This
crossover is clearly seen in Fig.\ref{coll}.
In accordance with the above observations we propose the
following scaling functions for the whole range of $\epsilon$.
Defining $u=l/t^{1/z}$ and
$s=y_L/l^{\chi}$, we have,
\begin{eqnarray}
C_h &\sim &\alpha_h(\ep)\; t^\beta \;f(u,s)\cr
C_x &\sim &\alpha_x(\ep)\; t^{1/z} \;g(u,s)\end{eqnarray}
where 
\be f(u,s)\sim \cases{ {\rm const.} &$ u\gg 1$\cr
u^\chi  &$ u\ll 1,\; s\ll 1$\cr
us^{1/z} & $u\ll 1,\; s\gg 1$}
 \label{summf}\ee
and
\be g(u,s)\sim \cases{ u
&$\,\,\,\,\,\,\,\,\,\,\,\,\,\,\,\,\,\,\,\,
s\ll 1 $\cr
u^{\chi}/s &$u\gg 1,\; s\gg 1$\cr
s^{-1/\tilde\beta z} &$u\ll 1,\; s\gg 1.$}\label{summg}\ee
where $\chi,z,$ and $\beta$ have their KPZ values.
The amplitudes are defined as $\alpha_x(\ep) =
(a+\aep^{1/6})$ and $\alpha_h=1/(a^{-1}+\aep^{1/6})$, where $a$
is some constant.
\begin{figure}
\begin{center}
\leavevmode
\vspace*{0.3cm}
\psfig{figure=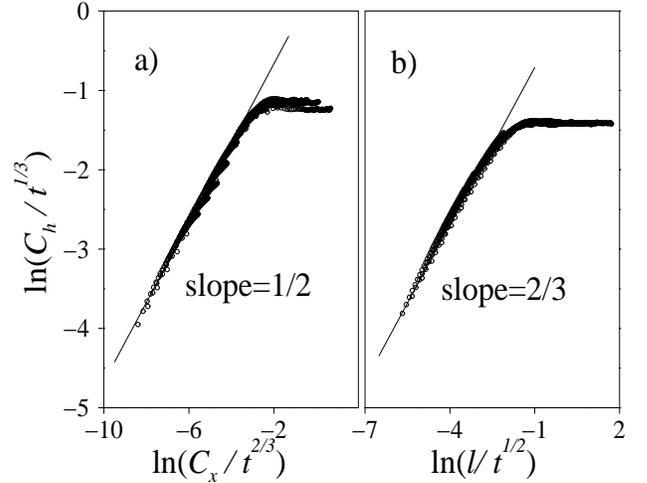,width=8cm,height=6cm,angle=0}
\end{center}
\narrowtext
\caption{a) $C_h$ and $C_x$ for $\epsilon=0.5$ for
$l=2,...,L/2$, and different times. b) $C_h$ in the isotropic
region,
$\epsilon =10^{-4}$.}
\label{chcx}
\end{figure}
From these scaling forms and Eq.(\ref{yL}) we see that 
the new critical exponents obey the relationships,
\be
\psi=\tilde\beta -{1 \over\zk} +{\chik \over \zk}
\label{psii}
\ee
and
 \be 
\chiis=\zeta \chik/\zk 
\label{chii}
\ee
which yields,
\be \chiis =\betak/\tilde\beta\;\;\;.\label{tildeb}\ee

From Eqs.(\ref{CxE2},\ref{Chc2}) we see 
that the graph dimension of a singly connected part in the
isotropic regime is  $D_g=1/\chiis$. Since in two dimensions,
$D_g$ is related to the roughness exponent via
 $D_g=2-\chi$, for $\chi=1/2$ we get
$\chiis=2/3$. The scaling relation (\ref{tildeb}) yields
$\tilde\beta=1/2$, from (\ref{psii}) and (\ref{chii}) it follows that
$\psi=1/6$
and
$\zeta=2$.
The fractal dimension of the self-similar set of interface
sites\cite{son} within a band of width  $y_L \sim \aep^{-\nu}$
is found, from boxcounting to be $D_I=1.85 \pm 0.05$  for length
scales
$\ell<y_L^2$.

In conclusion, we have presented an absorption-reaction  
model where the interface undergoes a 
delocalisation transition at the point where the mean velocity
of the 
interface goes to zero.
Although it might be conjectured~\cite{Tim} that as the velocity
of the interface vanishes, the scaling behaviour should cross
over to the Edwards-Wilkinson universality class, this is not the
case here. 
It has previously been observed~\cite{Nolle,Cieplak}, 
that the presence 
of overhangs, islands and inclusions may cause
the small-scale structure of the interface to crossover form
being self-affine to self-similar 
while the large scale behaviour remains
self-affine.  In the present model, this crossover is driven by
a competing length scale, the thickness of the interface, which
diverges at the critical point as $\aep^{-\nu}$ with $\nu=1/2$.
In the critical region, 
the interface is characterized by new set of
exponents $\chi^{\rm isot}=2/3$, $\tilde\beta=1/2$, $\zeta=2$, 
$\psi=1/6$, and
the fractal dimension $D_I=1.85$.  Except for $\nu$ and $D_I$,
these exponents may be obtained from the KPZ exponents via
scaling relations.

It should finally be mentioned that the reaction region can be
described by stochastic differential equations of the
multiplicative noise type with a single component field, since
in our model interface sites cannot be created spontaneously
either in the bulk or the vacant region. A field-theoretic 
renormalization group computation a la 
Tu, Grinstein and Mu\~noz~\cite{Grinstein} is presently under way 
to obtain the values of the critical exponents.

{\bf Acknowledgements}
A.K. would like to thank the G\"ursey Institute, where this work
was initiated, for their hospitality and gratefully acknowledges
support
by the Scientific and Technical Research Council of Turkey
(T\"UBITAK)
and by the U.S. National Science Foundation under Grant No.
DMR-94-00334.
A.E. thanks Mustansir Barma, Satya Majumdar, Deepak
Dhar, and Sondan Durukano\u glu Feyiz, for  a number of useful
conversations 
and acknowledges partial support from the Turkish Academy of
Sciences.

\end{multicols}
\end{document}